\documentclass[pra,a4paper,twocolumn,superscriptaddress,longbibliography,nofootinbib]{revtex4-2}
\usepackage[utf8]{inputenc}
\usepackage{amssymb}
\usepackage{amsmath}
\usepackage{amsfonts}
\usepackage{amsthm}
\usepackage{amsbsy}
\usepackage{bm}
\usepackage{bbold}

\usepackage{graphicx}
\usepackage{xcolor}
\usepackage{multirow}
\usepackage{float}
\usepackage{comment}
\usepackage{ulem}
\usepackage{relsize}
\usepackage{cmap}
\usepackage[colorlinks]{hyperref}

\DeclareMathOperator{\erf}{erf}
\DeclareMathOperator{\erfc}{erfc}

\begin{document}
	
	\title{Comment on ``Super-universality in Anderson localization''}
	
	\author{I. S. Burmistrov}
	
	\affiliation{\hbox{L.~D.~Landau Institute for Theoretical Physics, acad. Semenova av. 1-a, 142432 Chernogolovka, Russia}}

	\date{\today} 
	
\begin{abstract}

\end{abstract}

	\maketitle
	
Recently, Ref. \cite{Horvath2022} investigated the quantity $\mathcal{N}_*$, termed there ``minimal effective amount'' or ``minimal counting scheme'', at the Anderson transitions in orthogonal (AI), unitary (A), symplectic (AII), and chiral unitary  (AIII)  classes in three spatial dimensions, $d{=}3$.
This quantity, specifying an effective volume of the wave-function support, is defined on a lattice by
\begin{equation}
\mathcal{N}_*  =  \sum_{j=1}^N \min\{N |\psi(\bm{r}_j)|^2, 1\},  
\label{eq:N:def}
\end{equation}
where $\psi(\bm{r}_j)$ is a wave function on a site with coordinate $\bm{r}_j$. 
The authors of Ref. \cite{Horvath2022} presented numerical evidence for a ``super-universal'' (intact for all four symmetry classes studied) power-law scaling of the quantity \eqref{eq:N:def} averaged over disorder realizations, 
\begin{equation}
\langle \mathcal{N}_* \rangle\sim L^{d_{{\rm IR}}}  \quad {\rm with} \quad  d_{{\rm IR}}\approx 8/3.
\label{eq:wrong} 
\end{equation}

Below, for the standard Wigner-Dyson (WD) symmetry classes, we shall demonstrate that: 
\begin{itemize}
\item[(i)] the quantity $\langle \mathcal{N}_* \rangle$ is naturally related to the singularity spectrum function $f(\alpha)$ characterizing multifractality at Anderson transitions and  $d_{\rm IR}$ is nothing but $f(d)$; 
\item[(ii)] 
the scaling of $\langle \mathcal{N}_* \rangle$ with the system size $L$ is not of purely power-law character but rather contains an additional logarithmic factor.

\item[(iii)] $\langle \mathcal{N}_* \rangle$ does not demonstrate ``super-universality'' in a strict sense: its behavior does depend on the symmetry class. 
\end{itemize}

Let us consider a random quantity $\alpha$, which is related with a wave function amplitude as $|\psi(\bm{r}_j)|^2\sim L^{{-}\alpha}$. Introducing its distribution function $\mathcal{P}(\alpha)$ with the normalization condition, 
\begin{equation}
\int d\alpha \, \mathcal{P}(\alpha) = N \equiv L^d ,
\label{eq:norm}
\end{equation}
we rewrite the averaged quantity \eqref{eq:N:def} as
\begin{equation}
\langle \mathcal{N}_* \rangle =  \int d\alpha\, \mathcal{P}(\alpha) \min\{L^{d-\alpha},1\} .
\label{eq:N:def2}
\end{equation}
We note that, in the absence of randomness, $\mathcal{P}(\alpha)=L^d\delta(\alpha-d)$ such that $\langle \mathcal{N}_* \rangle\equiv L^d$.

In the Anderson localization problem, the distribution function $\mathcal{P}(\alpha)$ is expressed in terms of the so-called singularity spectrum function $f(\alpha)$ 
(see Ref.~\cite{Evers2008} and references therein):
\begin{equation}
\mathcal{P}(\alpha) = \mathcal{C} L^{f(\alpha)},
\label{eq:f:def}
\end{equation}
where $\mathcal{C}$ is the normalization constant.
Importantly, Eqs. \eqref{eq:N:def2} and \eqref{eq:f:def} indicate that the minimal effective amount $\langle \mathcal{N}_* \rangle$ is fully determined by the singularity spectrum $f(\alpha)$.
The latter is a well-known and well-studied object characterizing multifractal properties at Anderson transitions. The physical meaning of the function $f(\alpha)$ is nicely described in Ref.~\cite{Evers2008} as the fractal dimension of the set of those points $\mathbf{r}$ where the eigenfunction intensity is $|\psi^2(\mathbf{r})|\sim L^{-\alpha}$, which makes a direct connection to Eq. (\ref{eq:N:def}).
The Lagrange transform of $f(\alpha)$ gives the multifractal spectrum $\tau_q$. We note that the form of the singularity 
spectrum depends on the symmetry class and spatial dimensionality. Thus, Eqs. \eqref{eq:N:def2} and  \eqref{eq:f:def} suggest a dependence of $\langle \mathcal{N}_* \rangle$ on the symmetry class. 

We remind the reader that $f(\alpha)$ has negative second derivative, $f^{\prime\prime}(\alpha){<}0$, and a maximum at the point $\alpha_0{>}d$
with magnitude  $f(\alpha_0){=}d$. For Anderson localization problem in WD classes, the function $f(\alpha)$ satisfies the symmetry relation \cite{Evers2008},
\begin{equation}
f(2d-\alpha)=f(\alpha)+d-\alpha . 
\label{eq:sym}
\end{equation}

In the thermodynamic limit $L{\to}\infty$, the normalization constant $\mathcal{C}$ in Eq.~(\ref{eq:f:def}) is determined by the range of $\alpha$ in the vicinity of the maximum of $f(\alpha)$. Thus, we obtain the following estimate:
\begin{gather}
\mathcal{C} = c_0 \sqrt{\ln L} ,\qquad c_0 = \sqrt{|f^{\prime\prime}(\alpha_0)|/(2\pi)} .
\label{C-and-c0}
\end{gather}
Using relation \eqref{eq:sym}, we rewrite the integral in Eq. \eqref{eq:N:def2} as 
\begin{gather}
\langle \mathcal{N}_* \rangle = 2 \mathcal{C} \int\limits_{-\infty}^d d\alpha \, L^{f(\alpha)} .
\label{eq:integral}
\end{gather}	
Provided the 
condition $\ln L \gg |f^{\prime\prime}(d)|$ holds, the integral over $\alpha$ in Eq.~(\ref{eq:integral}) is dominated by vicinity of the end-point, $\alpha{=}d$, so that we obtain
\begin{gather}
\langle \mathcal{N}_* \rangle \simeq 2 \mathcal{C} L^{f(d)} \int\limits_{-\infty}^d d\alpha L^{(\alpha-d)/2} = 4 c_0 \frac{L^{f(d)}}{\sqrt{\ln L}} .
\label{eq:main}
\end{gather}
Here, we used the exact relation $f^\prime(d){=}1/2$ that follows from the symmetry condition \eqref{eq:sym}. We emphasize that $\langle \mathcal{N}_* \rangle$ is expressed in terms of the two symmetry-class specific quantities, $f(d)$ and $c_0$. Thus,  Eq. \eqref{eq:main} proves the absence of ``super-universality'' of $\langle \mathcal{N}_* \rangle$.

\begin{table}[ht]
\caption{$f(\alpha{=}d{=}3)$ and $c_0$  at Anderson transitions in the WD classes in three dimensions. In order to estimate $c_0$ we used data from Ref.~\cite{Ujfalusi}.}
\begin{tabular}{c|c|c|c}
 & class AI & class A & class AII \\
\hline 
\hline
$f(3)$  & $2.730\div2.736$~\cite{Ujfalusi} & $2.719\div2.721$~\cite{Ujfalusi} & $2.712\div2.715$~\cite{Ujfalusi} \\
& $2.7307\div2.7328$~\cite{Rodriguez} & $2.7187\div2.7195$~\cite{Lindinger} &  \\
\hline
$c_0$ & 0.291  & 0.282   & 0.278  
\end{tabular}
\label{Tab}
\end{table}
\begin{figure}[t]
\centerline{\includegraphics[width=0.9\columnwidth]{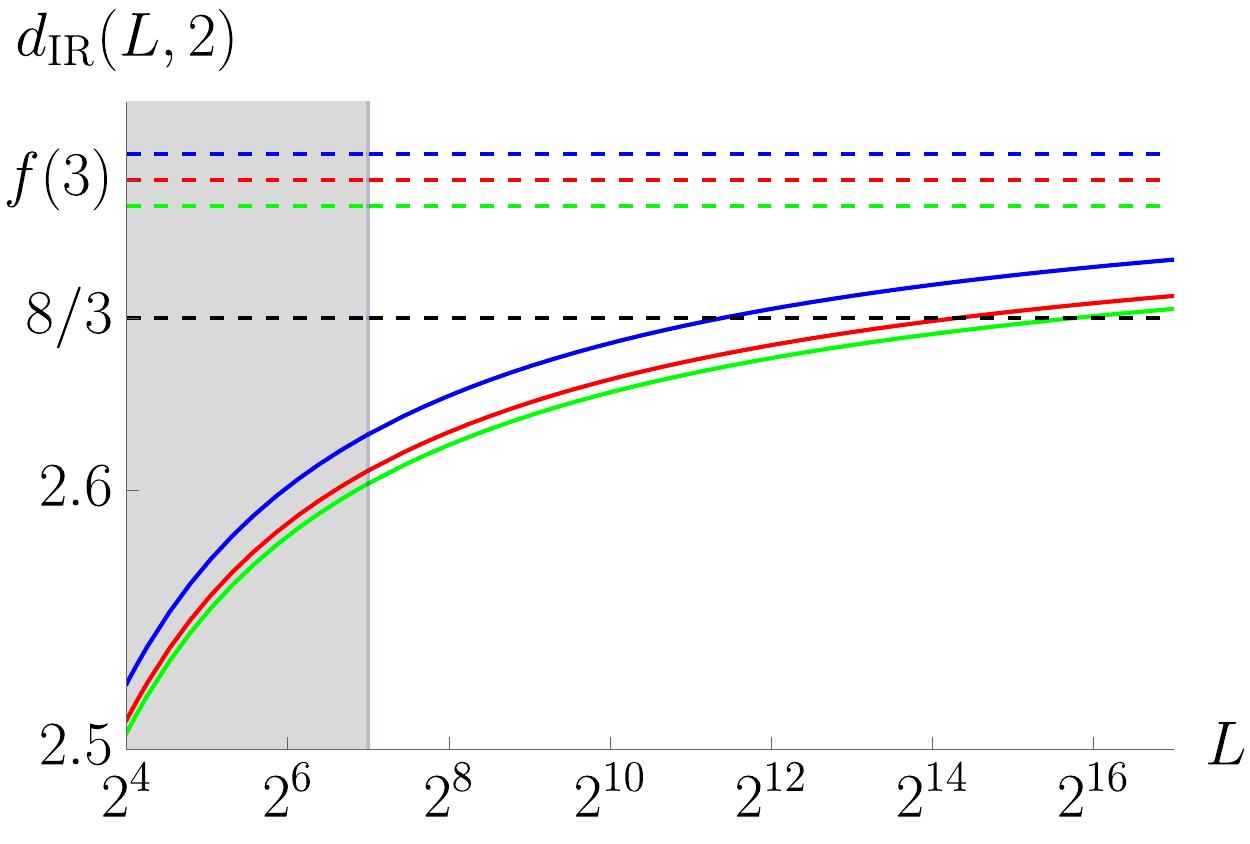}
}
\caption{Plot of 
$d_{\rm IR}(L,s)$ 
defined by Eq.~(\ref{eq:dir:def}),
as a function of $L$ on the logarithmic scale for $s=2$. Blue, red, and green curves correspond to 
Eq.~ \eqref{eq:main} for the symmetry classes AI (orthogonal), A (unitary), and AII (symplectic), respectively. 
The parameters of the curves are taken from Table \ref{Tab}.
The limiting value $8/3$ for $d_{\rm IR}$
proposed in Ref.~\cite{Horvath2022} is shown by the black dashed line. The blue, red, and green dashed lines indicate the asymptotic expression $f(d{=}3)$ for $d_{\rm IR}(L,2)$ in the limit $L{\to}\infty$ for the classes AI, A, and AII, respectively
The shaded area denotes the region of system sizes, for which numerical simulations in Ref. \cite{Horvath2022} were performed.
}
\label{fig}
\end{figure}

The Anderson transition in $d{=}3$ takes place in the strong-coupling limit, where controlled analytical calculations of the numerical values of relevant quantities are typically impossible. Singularity spectrum functions for $d{=}3$ Anderson transitions in standard WD classes have been computed numerically with high precision in Refs. \cite{Rodriguez,Ujfalusi,Lindinger}. 
As shown in Table \ref{Tab}, the numerical results give certainly distinct values of $f(\alpha{=}d{=}3)$ (as well as of $c_0$) in the three standard WD classes at $d{=}3$ Anderson transitions. 

It is worth noting a striking numerical closeness of the values of $f(d{=}3)$, which might indeed suggest a kind of universality, as hypothesized by the authors of Ref.~\cite{Horvath2022}.  Moreover, the whole singularity spectrum functions are very close (albeit certainly distinct) in $d{=3}$ for classes A, AI, and AII \cite{Ujfalusi}. However, this fact is specific for Anderson transitions in $d{=}3$. Indeed, in the case of Anderson transition in $d{=}2{+}\epsilon$ dimensions, one finds
$f(d){\simeq} d {-} b \epsilon/16$, where $b{=}4$ and $1$ for the classes AI and A, respectively (see, e.g. Ref.~\cite{Evers2008}). Therefore, in $d{=}2{+}\epsilon$ dimensions
$\langle \mathcal{N}_* \rangle$ clearly demonstrates no ``super-universality''. For the class AII situation is even more interesting, since Anderson transition occurs already in $d{=}2$ dimensions and $f(d{=}2){\simeq}2{-}0.04$~\cite{Mildenberger}. Thus, there is no reason to expect exact ``super-universality'' in $d{=}3$, either.

It is also worthwhile emphasizing that, according to Eq. \eqref{eq:main}, the scaling of $\langle \mathcal{N}_* \rangle$ with the system size is not purely power-law like, in contrast to the assumption of Ref.~\cite{Horvath2022}. The presence of $\sqrt{\ln L}$ in the denominator affects significantly analysis of the $L$ dependence at not too large $L$, see Fig.~\ref{fig}, where we plotted the function
\begin{equation}
d_{\rm IR}(L,s)=\frac{1}{\ln s}\,\ln \frac{\langle \mathcal{N}_*(L) \rangle}{\langle \mathcal{N}_*(L/s) \rangle},
\label{eq:dir:def}
\end{equation}
introduced in Ref.~ \cite{Horvath2022}, as a function $L$ on the logarithmic scale for $s=2$.  This is the reason that the exponent $d_{{\rm IR}}$ found in Ref.~\cite{Horvath2022} by extrapolating the results for $L\leq 128$ to $L\to\infty$ is smaller than $f(d)$.

The result \eqref{eq:main} holds also for the typical value of $\mathcal{N}_*$. In order to compute it one needs to restrict the integral over $\alpha$ in Eq.~\eqref{eq:N:def2} to the interval $\alpha_-{<}\alpha{<}\alpha_+$ at which $f(\alpha){>}0$. Since the integral is dominated by vicinity of the point $\alpha{=}d$ which is well inside the above interval we find the same result \eqref{eq:main} for the typical value of $\mathcal{N}_*$.

Finally, we note that in definition of $\mathcal{N}_*$ instead of $d$ one could use any value of $\alpha_*$ in the range $\alpha_1{<}\alpha_*{<}\alpha_0$ where the special point $\alpha_1$ satisfies $f(\alpha_1){=}\alpha_1$ and $f^\prime(\alpha_1){=}1$. Then $L$ dependence of $\mathcal{N}_*$ would be given by Eq. \eqref{eq:main} with $\alpha_*$ instead of $d$. 

\vspace{0.8cm}
 
The author thanks Igor Gornyi as one of the authors of the original idea of this comment for his continuous interest and enormous contribution to this work.  The author thanks J. Dieplinger, F. Evers, and K. Slevin for useful correspondence. The author is especially grateful to A. Mirlin for useful comments.

\begin{figure}[b]
\centerline{\includegraphics[width=0.9\columnwidth]{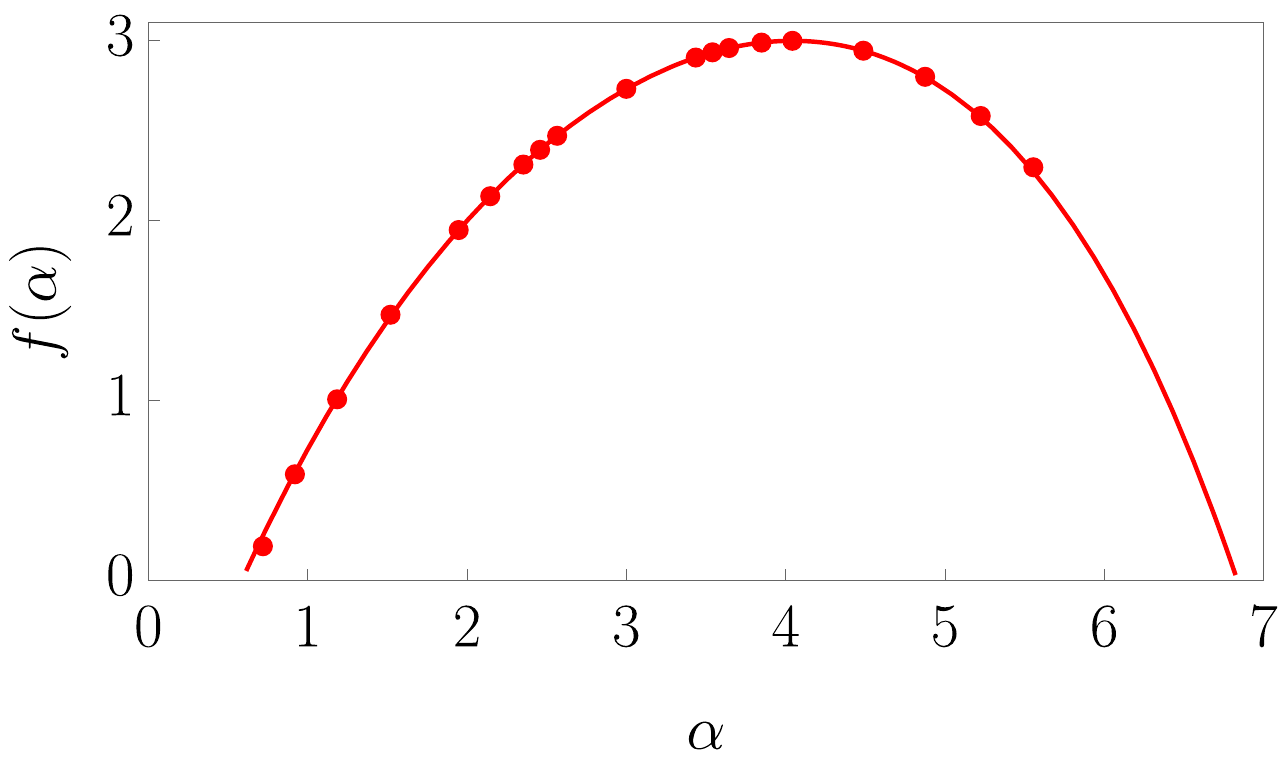}
}
\caption{The numerical data for $f(\alpha)$ in the class AI for $d{=}3$ from Ref. \cite{Ujfalusi} and the fit in accordance with Eq. \eqref{eq:f:fit}.
}
\label{fig:f:Fit}
\end{figure}
 
\vspace{0.5cm}

\centerline{\large \textsf{ADDENDUM I}}

\vspace{0.5cm}

\begin{figure*}[t]
\centerline{\includegraphics[width=0.9\columnwidth]{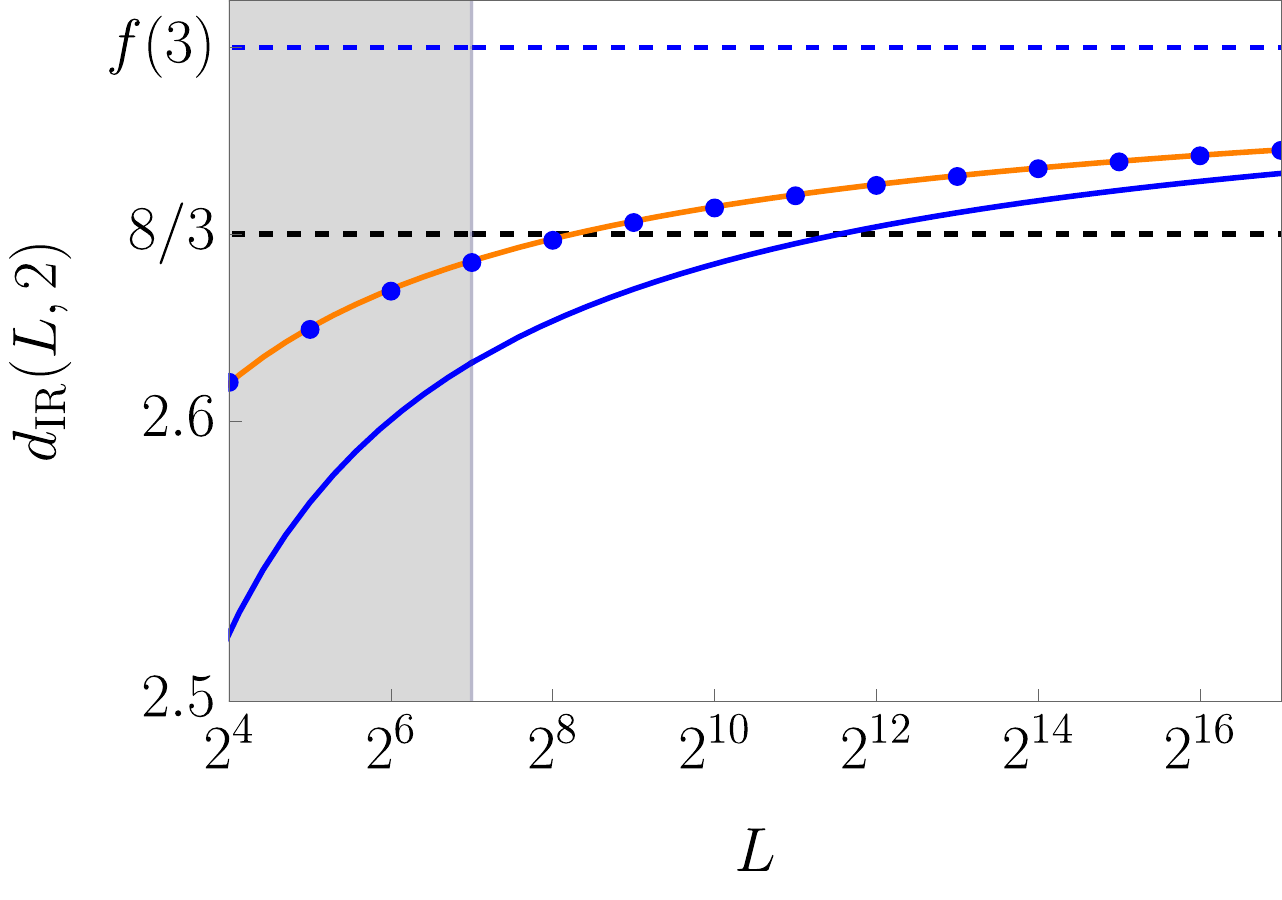}\qquad \qquad
\includegraphics[width=0.9\columnwidth]{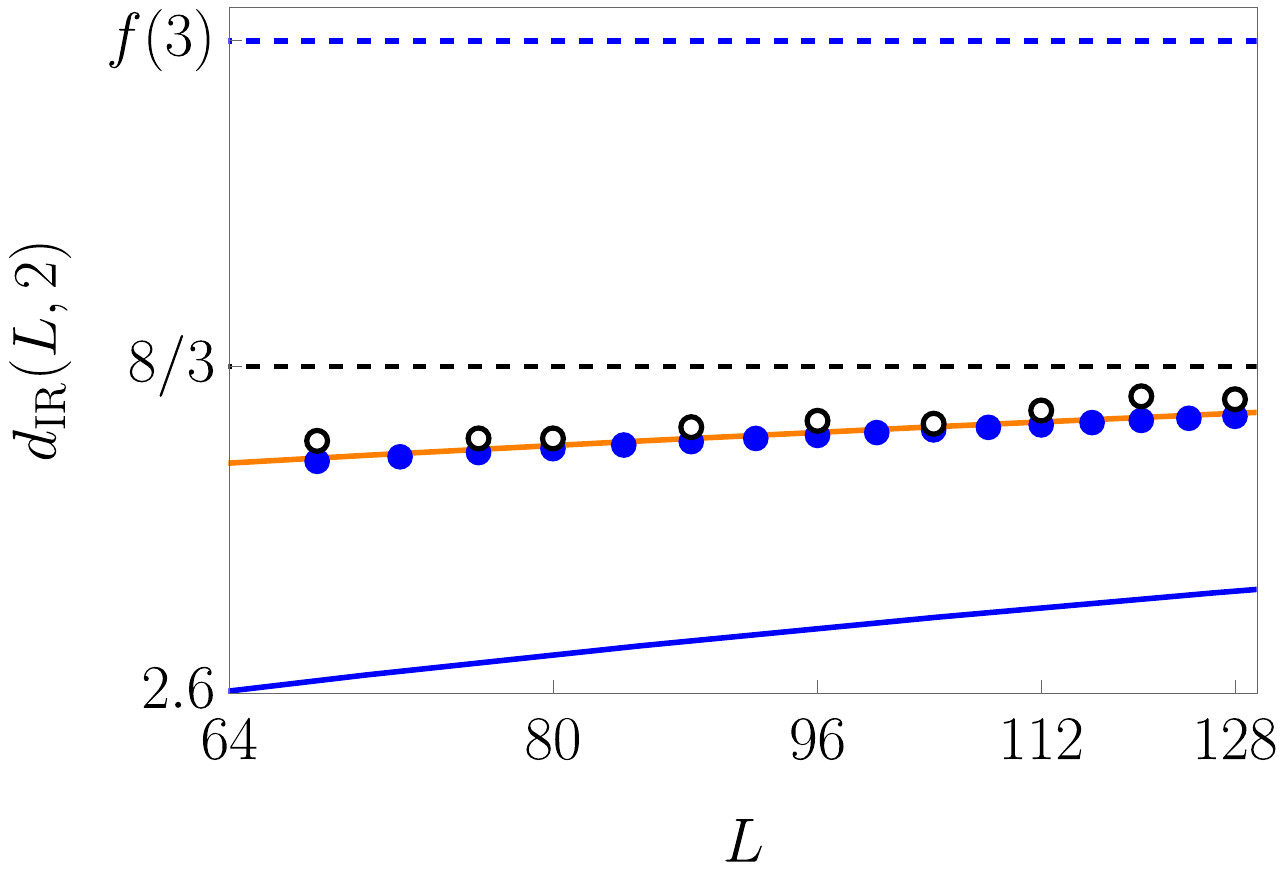}
}
\caption{Plot of 
$d_{\rm IR}(L,2)$ (for class AI)
defined by Eq.~\eqref{eq:dir:def},
as a function of $L$ on the logarithmic scale. Blue dots are numerical data for 
$d_{\rm IR}(L,2)$ using Eqs. \eqref{eq:integral} and \eqref{eq:dir:def} with 
the fitted singularity spectrum \eqref{eq:f:fit}. The blue solid curve is the analytic approximation corresponding to Eq.~\eqref{eq:main}, yielding the asymptotics of $d_{\rm IR}(L,2)$ in the thermodynamic limit. The orange solid curve is an analytic result based on parabolic approximation for $f(\alpha)$, see Eq.~\eqref{eq:main:par}. 
The limiting value $8/3$ for $d_{\rm IR}$
proposed in Ref.~\cite{Horvath2022} is shown by the black dashed line. The blue dashed line indicates the asymptotic expression $f(3)$ for $d_{\rm IR}(L,2)$ in the limit $L{\to}\infty$. The shaded area denotes the range of system sizes, for which numerical simulations in Ref. \cite{Horvath2022} were performed. 
The right panel is an enlarged part of the shaded region from the left panel. Empty black dots are data obtained in numerical simulations of Ref.~\cite{Horvath2022}.
}
\label{fig:2}
\end{figure*}

The material below has been added after Response to the comment has appeared \cite{Reply}, where the basics of the multifractality theory of Anderson transitions \cite{Evers2008} were questioned, in particular, Eq.~(\ref{eq:f:def}).
Here, we verify the full expression \eqref{eq:integral} for $\langle \mathcal{N}_* \rangle$ following from Eq.~(\ref{eq:f:def}) and compare the result with the $L{\to} \infty$ asymptotic expression \eqref{eq:main},
as well as with the available numerical data.

Using the numerical data of Ref.~\cite{Ujfalusi} we extract the following approximate form for the
singularity spectrum $f(\alpha)$ in the orthogonal symmetry class AI for $d{=}3$:  
\begin{gather}
f(\alpha)\simeq 3 - 0.2662 (\alpha-\alpha_0)^2 - 
 0.0254 (\alpha-\alpha_0)^3 \notag \\
 - 
 0.0061 (\alpha-\alpha_0)^4,
 \label{eq:f:fit}
\end{gather}
where $\alpha_0{\approx} 4.043$. 
We note that the approximate function \eqref{eq:f:fit} satisfies the symmetry relation, Eq.~\eqref{eq:sym} (maximum deviation of $f(6{-}\alpha){-}f(\alpha){+}\alpha{-}3$ from zero is 0.006; the deviation can be further reduced by keeping more digits in the coefficients). 
The comparison between numerical data from Ref.~\cite{Ujfalusi} and Eq.~\eqref{eq:f:fit} is shown in Fig. \ref{fig:f:Fit}. As a check of quality of the fit we compute the difference $f^\prime(3){-}1/2$ that is equal to $0.8\cdot 10^{-4}$ 
instead of zero. The fitting function \eqref{eq:f:fit} displays a weak deviation from exact parabolicity, in a full agreement with the conclusions of Refs.~\cite{Ujfalusi}
and \cite{Rodriguez}.

Next, we use the fitted singularity spectrum \eqref{eq:f:fit} to compute $d_{\rm IR}(L,2)$ numerically following Eqs.~\eqref{eq:integral} and \eqref{eq:dir:def}.
The result is shown in Fig.~\ref{fig:2} by blue dots. As one can see, for the region of system sizes for which numerical simulations in Ref.~\cite{Horvath2022} were performed, evaluation of the full expression \eqref{eq:integral} yields larger values of $d_{\rm IR}(L,2)$ than those given by asymptotic result \eqref{eq:main} (blue solid line).

We note that approximating $f(\alpha)$ by a parabola, i.e., omitting the cubic and quartic terms in Eq.~\eqref{eq:f:fit}, one can compute the integral determining $\mathcal{N}_*$ in general formula \eqref{eq:integral} exactly, yielding a compact analytical expression:
\begin{equation}
\langle \mathcal{N}_* \rangle \simeq 4 c_0 \frac{L^{f(d)}}{\sqrt{\ln L}} \Phi\left (\frac{\sqrt{\ln L}}{2\sqrt{2|f''(d)|}}\right ).
\label{eq:main:par}
\end{equation}
Here the function $\Phi(x){=}\sqrt{\pi} x \exp(x^2) \erfc(x)$, where $\erfc(x)$ is the complementary error function, tends to the unity as $x{\to}\infty$. Thus,
Eq.~(\ref{eq:main:par}) asymptotically reproduces Eq.~(\ref{eq:main}) in the limit ${L{\to}\infty}$. We show $d_{\rm IR}(L,2)$ obtained with the help of Eq.~\eqref{eq:main:par} by the orange curve in Fig. \ref{fig:2}. As one can see, the analytical result based on the parabolic approximation for $f(\alpha)$ nicely matches the result of numerical computation with the non-parabolic singularity spectrum function, Eq. \eqref{eq:f:fit}. It can be indeed expected in view of the numerical smallness of coefficients in front of the qubic and quartic terms in the expansion of $f(\alpha)$ near the maximum (weak non-parabolicity), see Eq.~\eqref{eq:f:fit}.  

Further, in the right panel of Fig. \ref{fig:2}, we focus on the region of system sizes, for which numerical simulations in Ref.~\cite{Horvath2022} were performed. Using Fig.~2 of Ref. \cite{Horvath2022}, we extract their numerical data for $d_{\rm IR}(L,2)$
and show them by empty black dots. There is almost perfect matching of the data of Ref.~\cite{Horvath2022} (empty black dots) with the full blue dots obtained numerically from our exact analytical expressions \eqref{eq:integral} and \eqref{eq:dir:def} with the help of the singularity spectrum function, Eq.~\eqref{eq:f:fit}, extracted from numerical simulations in Ref.~\cite{Ujfalusi}.

This fully confirms (i) the quality of the numerical data of Refs.~\cite{Ujfalusi} and \cite{Horvath2022} and (ii) the consistency of the data from these numerical studies with the prediction of the multifractality theory of Anderson transition. This also indicates that the range of system sizes considered in the above numerical works is indeed already sufficient to systematically capture the multifractal properties of the critical wave function with high precision. The parabolic approximation of the singularity spectrum (orange solid line) yields a remarkably good agreement with the data from Ref.~\cite{Horvath2022}, with small deviations that can be attributed to the effect of weak non-parabolicity. It thus turns out that the exponent $d_{\rm IR}(L,2)$ can be, in principle, used as an alternative indicator of non-parabolicity of the multifractal singularity spectrum. 

Finally, according to the definition \eqref{eq:dir:def}, the symmetry-class specific coefficient $c_0$, Eq.~(\ref{C-and-c0}), drops out from  $d_{\rm IR}(L,2)$. At the same time, the value of $\langle \mathcal{N}_* \rangle$ does depend on $c_0$, as well as on the ultraviolet scale, in units of which the system size is measured.
In the formulas above, the length scale $L$ is given in units of the lattice constant of the discretized model. In fact, the ultraviolet scale for the power-law multifractal scaling (\ref{eq:f:def}) is provided by the effective mean-free path, which, in contrast to the universal (for a given symmetry class) coefficient $c_0$, depends on the microscopic details (type of lattice, disorder distribution). At Anderson transitions in $d=3$ dimensions, the mean-free path $l_\text{mfp}$ is of the order of the lattice constant (as the critical conductance is of order unity) but may differ from it by a numerical factor of order 1. It can be easily checked that dividing $L$ in Eq.~(\ref{eq:main:par}) by 1.21, one can perfectly fit the curve for $\langle \mathcal{N}_* \rangle$ from Ref.~\cite{Horvath2022} in the full range of $L$ studied there, with the value $c_0=0.291$ (see Table I) extracted from the multifractal analysis. This simply means that for the microscopic model of Ref.~\cite{Horvath2022} the effective mean-free path is related to the lattice constant $a$ as $l_\text{mfp}\approx 1.21 a$, in a full agreement with theoretical expectations.
Note that the function $d_{\rm IR}(L,2)$ remains almost unaffected by such a change of the length units.

In conclusion, this addendum demonstrates a remarkable agreement between the numerical results of Ref.~\cite{Horvath2022} and the predictions of the multifractality theory of Anderson transitions, Eqs. (\ref{eq:f:def}) and (\ref{eq:integral}).

\vspace{1cm}

\centerline{\large \textsf{ADDENDUM II}}

\vspace{0.5cm} 

The material below has been added after the version V2 of the Response to the Comment has appeared  \cite{Reply}.

\vspace{0.5cm} 

\noindent \textbf{\underline{REMARK 1:} ``Moment-based'' versus ``genuine'' multifractality.}

\vspace{.2cm}

In Ref.~\cite{Reply}, a possible distinction between genuine multifractality (MF) and ``moment multifractality'' (mMF) was put forward. This conjecture was motivated by the consideration of a set of points where the intensity of a critical wave function satisfies a given condition on its magnitude. Indeed, in general, it is not obvious that such a definition of the multifractal dimensions is equivalent (at all scales, not only in the thermodynamic limit) to the one based on the analysis of the scaling of the moments of the wave-function intensity. A famous example of non-unique restoration of a distribution function from its moments is Stieltjes example for the log-normal distribution (see, e.g., Ref.~\cite{Schmudgen}). We note, however, that such an ambiguity arises only for the moments with integer powers $q$. 

We stress that, in the theory of multifractality of Anderson transitions \cite{Evers2008}, the moments of the wave-function intensities correspond to scaling eigenoperators of the underlying field theory \cite{Mirlin2000,Gruzberg2011, Gruzberg2013}. Therefore, these moments demonstrate a pure scaling behavior, i.e., are described by exact power laws of the system size, without any corrections to scaling,
\begin{equation}
    P_q(L)=\int d^d\bm{r}\, \langle |\psi|^{2q}\rangle= L^{-\tau_q}.
    \label{PqL}
\end{equation}
This statement holds true for an arbitrary power $q$ for the moments of the wave function at any single point in space \cite{Evers2008}. The condition of a strictly power-law scaling of these moments imposes very strict constraints on a possible form of the distribution function of wave-function intensities. It is also worth mentioning in passing that, for multi-point correlation functions, one can always choose proper combinations (expressed through Slater determinants) of wave functions, which are also described by a purely power-law scaling \cite{Gruzberg2011, Gruzberg2013}.      

As such, the moments of the wave-function intensity (as well as the moments of the local density of states) are extremely convenient objects for studying multifractality in numerical experiments on Anderson transitions. Numerically, multifractal exponents $\tau_q$ are computed not only at integer values of $q$ but at non-integer values of $q$ also (see, e.g., Ref.~\cite{Ujfalusi}). This fact allows one to extract $f(\alpha)$ from $\tau_q$ without ambiguity. We also remind the readers that the field-theory approach based on the nonlinear sigma model allows one to compute the tails of the distribution function $\mathcal{P}(\alpha)$ directly, avoiding computation of the moments \cite{Mirlin1996}. 

The singularity spectrum function $f(\alpha)$ (without ambiguity between mMF and MF) is well established both from the points of view of numerics and analytical theory. The equivalence of the two approaches to extracting the multifractal singularity spectrum was first shown long time ago in Ref.~\cite{Halsey1986}. In other words, the two definitions of the multifractal singularity spectrum, $f(\alpha)$ and $f_m(\alpha)$, as introduced in Ref.~\cite{Reply}, coincide. The whole spectrum $f(\alpha)$ can be obtained from the moments of the wave-function intensity in a standard way described already in Ref.~\cite{Halsey1986} [see Eqs.~(1.10) and (1.11) there].

\vspace{0.3cm}

\noindent \textbf{\underline{REMARK 2:} Saddle-point approximation: Appearance of logarithms.}

\vspace{0.2cm}

To highlight the difference between mMF and MF, the authors of Ref.~\cite{Reply} introduced the following form of the intensity distribution function (see Eqs.~(4)-(6) of Ref.~\cite{Reply}):
\begin{equation}
\mathcal{P}(\alpha,L)= v(\alpha,L) 
 L^{f(\alpha,L)},
\label{PHM}
\end{equation}
with the size-dependent function $f(\alpha,L)$ and $\alpha$-dependent weight $v(\alpha,L)$ of $\alpha$-populations. As discussed above, the equivalence of $f_m(\alpha)$ understood as the limit $L\to \infty$ of $f(\alpha,L)$ in Eq.~(\ref{PHM}) and $f(\alpha)$ restored by the Legendre transformation from the wave-function moments was demonstrated already in Ref.~\cite{Halsey1986}. Nevertheless, it is instructive to adopt the Ansatz (\ref{PHM}) of Ref.~\cite{Reply} to demonstrate how the logarithmic corrections to the ``canonical'' form of $\mathcal{P}(\alpha,L)$ [see Eq.~\eqref{eq:f:def}] may arise away from the asymptotic limit $L\to \infty$. While these corrections do not lead to corrections to the pure power-law scaling of the wave-function moments, Eq.~(\ref{PqL}), they can introduce additional logarithmic corrections to other--not purely scaling--observables, in particular, to $\langle \mathcal{N}_*\rangle$, which could be visible in simulations on numerically accessible system sizes.  
Although distribution \eqref{PHM} complicates analyses, the result for such observables in the asymptotic limit $L\to\infty$ is, of course, not sensitive to those complications. 

As rightly mentioned in Ref.~\cite{Reply}, there are two normalization conditions for the distribution function \eqref{PHM}:
\begin{align}
    \int_{-\infty}^\infty
    d\alpha\, \mathcal{P} (\alpha,L) L^{-d} &=1,
    \label{P1}
    \\
     \int_{-\infty}^\infty
    d\alpha\, \mathcal{P} (\alpha,L) L^{-\alpha} &=1. 
    \label{P2}
\end{align}
The condition \eqref{P1} is equivalent to Eq.~(\ref{eq:norm}) describing the total probability. The condition (\ref{P2}) describes the normalization of the wave functions. These two normalization conditions are, in fact, equivalent, owing to the symmetry property of the MF singularity spectrum, Eq.~\eqref{eq:sym}.

At first, we  calculate the moments of the wave-function intensity by employing the saddle-point approximation justified in the large-$L$ limit:
\begin{align}    P_q(L)&=\int d\alpha \mathcal{P}(\alpha,L) L^{-\alpha q}    = \int d\alpha\, e^{S_q(\alpha,L)},    
\\    
S_q(\alpha,L)&=\ln v(\alpha,L)+f(\alpha,L) \ln L - \alpha q \ln L,   
\label{eq18}
\\    
\frac{d S_q(\alpha,L)}{d\alpha}    &=\frac{v^\prime}{v}+f^\prime \ln L - q \ln L,    
\\    
\frac{d^2S_q(\alpha,L)}{d\alpha^2}    &=\frac{v^{\prime\prime}}{v}   -\frac{(v^\prime)^2}{v^2}+f^{\prime\prime} \ln L, 
\label{eq20}
\end{align}
where the prime denotes a derivative with respect to $\alpha$.
From the saddle-point condition, $S_q^\prime=0$, we obtain the relation
\begin{equation}
q=f^\prime(\alpha,L)
+\frac{v^\prime(\alpha,L)}{v(\alpha,L)\,\ln L}
\label{eq:qL:12}
\end{equation}
that determines the stationary-point value $\alpha=\bar{\alpha}_q(L)$.
This value depends on $L$ through the dependence of $f(\alpha,L)$ on $L$ and because of a nonzero value of $v^\prime(\alpha,L)$, i.e., the $\alpha$-dependence of $v(\alpha,L)$.
This saddle-point consideration clearly demonstrates how the logarithms of the the system size may appear in the problem. 

The saddle-point result for the $q$th moment (we remind the reader that here $q$ is not necessarily integer) of the wave-function intensity then reads:
\begin{align}
P_q(L)\approx \sqrt{\frac{\pi}{S_q^{\prime\prime}[\bar{\alpha}_q(L),L)]}}\ v[\bar{\alpha}_q(L),L]\  L^{f[\bar{\alpha}_q(L),L]-\bar{\alpha}_q(L) q}, 
\label{Pq-saddle}
\end{align}
where $S_q^{\prime\prime}[\bar{\alpha}_q(L),L)]$ is given by
Eq.~(\ref{eq20}) with functions $v$, $v^\prime$, $v^{\prime\prime}$ and $f^\prime$ evaluated at $\alpha=\bar{\alpha}_q(L)$.
A comparison of Eq.~(\ref{Pq-saddle}) and the exact formula \eqref{PqL}
shows that $v(\alpha,L)$ should be related to $f(\alpha,L)$ in an extremely strict manner, which goes beyond normalization conditions (\ref{P1}) and (\ref{P2}). Indeed, the exponent ${f[\bar{\alpha}_q(L),L]-\bar{\alpha}_q(L) q}$ and the prefactor in Eq.~(\ref{Pq-saddle}) should become $L$-independent: we reiterate that any corrections to a pure power-law scaling of the moments are strictly prohibited.

The normalization condition (\ref{P2}) is straightforwardly obtained by requiring $P_{q=1}(L)=1$.
Performing the same saddle-point analysis for the normalization integral in Eq.~(\ref{P1}), we find for the corresponding stationary-point value of $\alpha=\bar{\alpha}_0(L)$
\begin{equation}
    0=f^\prime(\bar{\alpha}_0,L)
+\frac{v^\prime(\bar{\alpha}_0,L)}{v(\bar{\alpha}_0,L)\,\ln L},
\label{eq:qL:12-1}
\end{equation}
resulting in
\begin{align}
1\approx \sqrt{\frac{\pi}{S_q^{\prime\prime}[\bar{\alpha}_0(L),L)]}}\ v[\bar{\alpha}_0(L),L]\  L^{f[\bar{\alpha}_0(L),L]-d} . 
\label{P1-saddle}
\end{align}
This saddle-point representation of the normalization condition again demonstrates a highly nontrivial relation between 
$v(\alpha,L)$ and $f(\alpha,L)$ defined by Ansatz (\ref{PHM}).

In addition, the deviations from exact parabolicity of the singularity spectrum (i.e., going beyond the saddle-point approximation) will introduce extra $L$-dependent corrections to the saddle-point results for the normalization integrals and wave-function moments. Indeed, this can be straighforwardly seen by expanding Eq.~(\ref{eq18}) beyond the parabolic order around $\bar{\alpha}_q(L)$ and representing the exponential of the terms with $S_q^{(3)}(\bar{\alpha}_q)(\alpha-\bar{\alpha}_q)^3$, $S_q^{(4)}(\bar{\alpha}_q)(\alpha-\bar{\alpha}_q)^4,\ \ldots$ in a series in the corresponding corrections. Each term in this series would then produce a power of $1/\ln L$ (again logarithmic corrections). It turns out that all such logarithmic corrections arising from different sources cancel out in the moments of the wave-function intensity, while accumulating in the asymptotic series in $1/\ln L$ in quantities that do not correspond to purely scaling operators of the field theory (like, e.g., $\langle \mathcal{N}_*\rangle$).

\vspace{0.3cm}

\noindent \textbf{\underline{REMARK 3:} 
Distibution function for wave-function intensities at finite scales $L$.}

\vspace{0.2cm}

Let us now demonstrate how the direct calculation of the wave-function moments is reconciled with their pure scaling form. Before going beyond the saddle-point approximation, it is instructive to consider the parabolic Ansatz for the singularity spectrum:
\begin{equation}
   f(\alpha)=d - \frac{(\alpha-\alpha_0)^2}{4(\alpha_0-d)}, \quad \mathcal{P}(\alpha,L) = \frac{\sqrt{\ln L}}{2\sqrt{\pi} \sqrt{(\alpha_0-d)}} L^{f(\alpha)} .
   \label{eq:f:p}
\end{equation}
Note that $f(\alpha)$ in Eq.~\eqref{eq:f:p} satisfies the symmetry relation \eqref{eq:sym}.
For this parabolic Ansatz, all the discussed quantities can be computed exactly. It is worth mentioning that the exact parabolic form of the multifractal spectrum is known to be exact for some non-Wigner-Dyson symmetry classes, e.g., for the chiral unitary class AIII~\cite{Evers2008}.

Using the simplicity of parabolic singularity spectrum, we explicitly compute the $q$th moment of the wave function,
\begin{gather}
P_q(L)=\int \limits_{-\infty}^\infty d\alpha \, L^{-\alpha q} \mathcal{P}(\alpha) = L^{-\tau_q}, \notag \\  \tau_q = d(q-1)-q(q-1)(\alpha_0-d).
\label{eq:tau}
\end{gather}
We see that this calculation is equivalent to the saddle-point calculation with $L$-independent $f(\alpha)$ and $\alpha$-independent $v(L)$. In particular, the prefactor in Eq.~(\ref{Pq-saddle}) becomes unity by construction: $\pi v/S_q^{\prime\prime}=1$ for Eq.~\eqref{eq:f:p}.

With the above example of a parabolic spectrum in mind, we observe that deviations from the parabolicity would lead to logarithmic correction to the moments if one used an $\alpha$-independent prefactor $\mathcal{C}=v(L)$ in the distribution function \eqref{eq:f:def}.
At the same time, Eq.~\eqref{eq:qL:12} suggests that $\alpha$-dependence of the prefactor $v(\alpha,L)$ would also yield the $1/\ln L$ corrections to the moments. Furthermore, based on Eq.~\eqref{eq:qL:12}, a similar slow $1/\ln L$ dependence could be expected to appear in $f(\alpha,L)$. However, because of the relation $L^{1/\ln L}=e$, one should always transfer such a slow $L$-dependence of $f(\alpha)$ into the pre-exponential factor $v(\alpha,L)$. Thus, $f(\alpha)\equiv f_m(\alpha)$. As we shall show below, the proper choice of $v(\alpha,L)$ guarantees the pure scaling form of wave function moments, Eq.~\eqref{PqL}, when going beyond the parabolic MF singularity spectrum. 

To illustrate the above statement, we write the pre-exponential function $v(\alpha,L)$ in the following form:
\begin{equation}
 v(\alpha,L)=\sqrt{\ln L} \,\sum_{j=0}^\infty \frac{c_j(\alpha)}{(\ln L)^j} .  
 \label{valphaL}
\end{equation}
One can then find the coefficients $c_j(\alpha)$ to be consistent with the pure scaling form of the wave function moments, as well as with the normalization conditions. The most general form of the MF distribution function is
\begin{equation}
    \mathcal{P}(\alpha,L) = v(\alpha,L) L^{f(\alpha)},
\label{eq:PalphaL-Halsey}
\end{equation}
where $v(\alpha,L)$ is completely determined by $f(\alpha)$.
This form of $\mathcal{P}(\alpha,L)$ was introduced already in Ref.~\cite{Halsey1986}.  

With Eq.~(\ref{valphaL}), we proceed by performing a regular expansion around the saddle-point solution for the moments of the intensity in the large-$L$ limit, $\ln L\gg 1$. 
In order to find coefficients $c_0$ and $c_1$ it is enough to consider terms of the order $1/\ln^2L$ (for finding further coefficients, one considers higher orders of inverse logarithm in a regular way). 
We get 
\begin{gather}
\tau_q \simeq \bar{\alpha}_q q - f^\prime(\bar{\alpha}_q) -\frac{1}{2\ln L} \ln \frac{2\pi c_0^2(\bar{\alpha}_q)}{|f^{\prime\prime}(\bar{\alpha}_q)|}
- \frac{c_1(\bar{\alpha}_q)}{\ln^2L} 
\notag\\
-\frac{1}{8\ln^2L}\Biggl 
(
\frac{f^{(4)}(\bar{\alpha}_q)}{[f^{\prime\prime}(\bar{\alpha}_q)]^2}
+\frac{5[f^{(3)}(\bar{\alpha}_q)]^2}{3[f^{\prime\prime}(\bar{\alpha}_q)]^2}
 -\frac{4[\ln c_0(\bar{\alpha}_q)]^{\prime\prime}}{[f^{\prime\prime}(\bar{\alpha}_q)]^2}  \Biggr ),
\end{gather}
where the derivatives are taken with respect to $\alpha$. The quantity $\bar{\alpha}_q$ is related with $q$ as follows:
\begin{gather}
q\simeq f^{\prime}(\bar{\alpha}_q)
+ \frac{[\ln c_0(\bar{\alpha}_q)]^{\prime}}{\ln L} + \frac{c_1^\prime(\bar{\alpha}_q)}{\ln^2 L} +\dots 
\end{gather}
Since $\bar{\alpha}_q$ depends on $L$, it is convenient to introduce $\alpha_q$ which is independent of $L$ and is related to $q$ in a standard way, $q=f^\prime(\alpha_q)$. Then we obtain
\begin{equation}
    \bar{\alpha}_q \simeq \alpha_q -\frac{1}{\ln L} \frac{[\ln c_0(\bar{\alpha}_q)]^{\prime}}{f^{\prime\prime}(\bar{\alpha}_q)}+\dots 
\end{equation}
The required purely power-law expression, Eq. \eqref{PqL}, with the exponent $\tau_q$ given by the conventional Legendre transform \cite{Halsey1986,Evers2008},
\begin{equation}
\tau_q=\alpha_q q - f(\alpha_q), \quad 
q=f^\prime(\alpha_q) ,
\end{equation}
is restored by the proper choice of the functions $c_0$, $c_1$, etc. at points $\alpha=\alpha_q$,
\begin{align}
    c_0(\alpha_q) & = \sqrt{\frac{|f^{\prime\prime}(\alpha_q)|}{2\pi}} ,\notag \\ 
    c_1(\alpha_q) & = \frac{1}{8} 
    \Biggl 
(
\frac{f^{(4)}(\alpha_q)}{[f^{\prime\prime}(\alpha_q)]^2}
-\frac{5[f^{(3)}(\alpha_q)]^2}{3[f^{\prime\prime}(\alpha_q)]^2}\Biggr ),\notag \\
\dots 
\label{eq:rel:gen}
\end{align}

Here, we recall a possible ambiguity in restoration of $\mathcal{P}(\alpha, L)$ from wave-function moments.  We restore the function $f(\alpha)$ and, consequently, functions $c_j(\alpha)$ at the points $\alpha=\alpha_q$.
Indeed, if the moments are available only at integer values of $q$, one can add arbitrary functions to $c_0(\alpha)$, $c_1(\alpha)$, etc., such that those functions are exactly zero at all $\alpha=\alpha_q$. These functions would not affect the scaling of the momenta but would modify the distribution function.  
However, as we mentioned in the Remark 1, such an ambiguity is resolved by allowing non-integer values of $q$ to make $\alpha_q$ essentially continuous variable, see Refs.~\cite{Ujfalusi} and \cite{Lindinger}. 

Thus, the pre-exponential function $v(\alpha,L)$ is indeed fully determined by the singularity spectrum function $f(\alpha)$, which can be restored from the exponents $\tau_q$. The multifractal singularity spectrum $f(\alpha)$ possesses the properties described in Sec. IIC2 of Ref.~\cite{Evers2008}. In particular, it satisfies the relations $f^\prime(\alpha_0)=0$ and $f(\alpha_0)=d$ as well as $f^\prime(\alpha_1)=1$ and $f(\alpha_1)=\alpha_1$, which guarantee the fulfillment of the normalization conditions \eqref{P1} and \eqref{P2} for $\mathcal{P}(\alpha,L)$. 
The explicit expressions above demonstrate that singularity spectrum $f(\alpha)$ extracted from $\alpha$-population coincides with singularity spectrum extracted  $f_m(\alpha)$ from the wave-function moments.

Finally, we note that the $\alpha$ dependence of the prefactor $v(\alpha,L)$ in Eq.~(\ref{valphaL}) results in $1/\ln L$ corrections to $\langle \mathcal{N}_*\rangle$:
\begin{align}
    \langle \mathcal{N}_*\rangle
    &=\frac{4 L^{f(d)}}{\sqrt{\ln L}}\sqrt{\frac{|f^{\prime\prime}(d)|}{2\pi}}
    \label{Nstar-logs}
    \\
    \times 
    \Biggl [ 
    1
    &+ \sqrt{\frac{2\pi}{|f^{\prime\prime}(d)|}}\left (\frac{f^{(4)}(d)}{8[f^{\prime\prime}(d)]^2}+4f^{\prime\prime}(d)\right )\frac{1}{\ln L} 
    + \dots
    \Biggr ].\notag 
\end{align}
In contrast to $P_q$ that demonstrates a pure power-law dependence on $L$, the quantity $\langle \mathcal{N}_*\rangle$ has corrections in the form of infinite series in $1/\ln L$, see also Eq.~\eqref{eq:main:par}. The origin of the logarithmic factors and terms here is straightforward. The overall $1/\sqrt{\ln L}$ factor trivially appears already at the level of the saddle-point approximation for the canonical distribution function (\ref{eq:f:def}), yielding the asymptotic expression (\ref{eq:main}). 
Further logarithmic corrections appear naturally from the two sources: (i) the integration in the general expression for $\langle \mathcal{N}_*\rangle$ is performed in finite limits and (ii) the deviation from the parabolicity is accounted for by going beyond the saddle-point approximation, as reflected in $\alpha$-dependence of $v(\alpha,L)$, Eq.~(\ref{valphaL}).
Needless to say, in the thermodynamic limit $L\to \infty$ (or, better to say, $\ln L \to \infty$), where the ``infrared dimension'' $d_\text{IR}$ is defined in Ref.~\cite{Horvath2022} as $d_\text{IR}=d_\text{IR}(L\to\infty)$,
these logarithmic corrections are absolutely immaterial for determination of the true value of $d_\text{IR}$.

\begin{figure*}[ht!]
\centerline{\includegraphics[width=0.95\columnwidth]{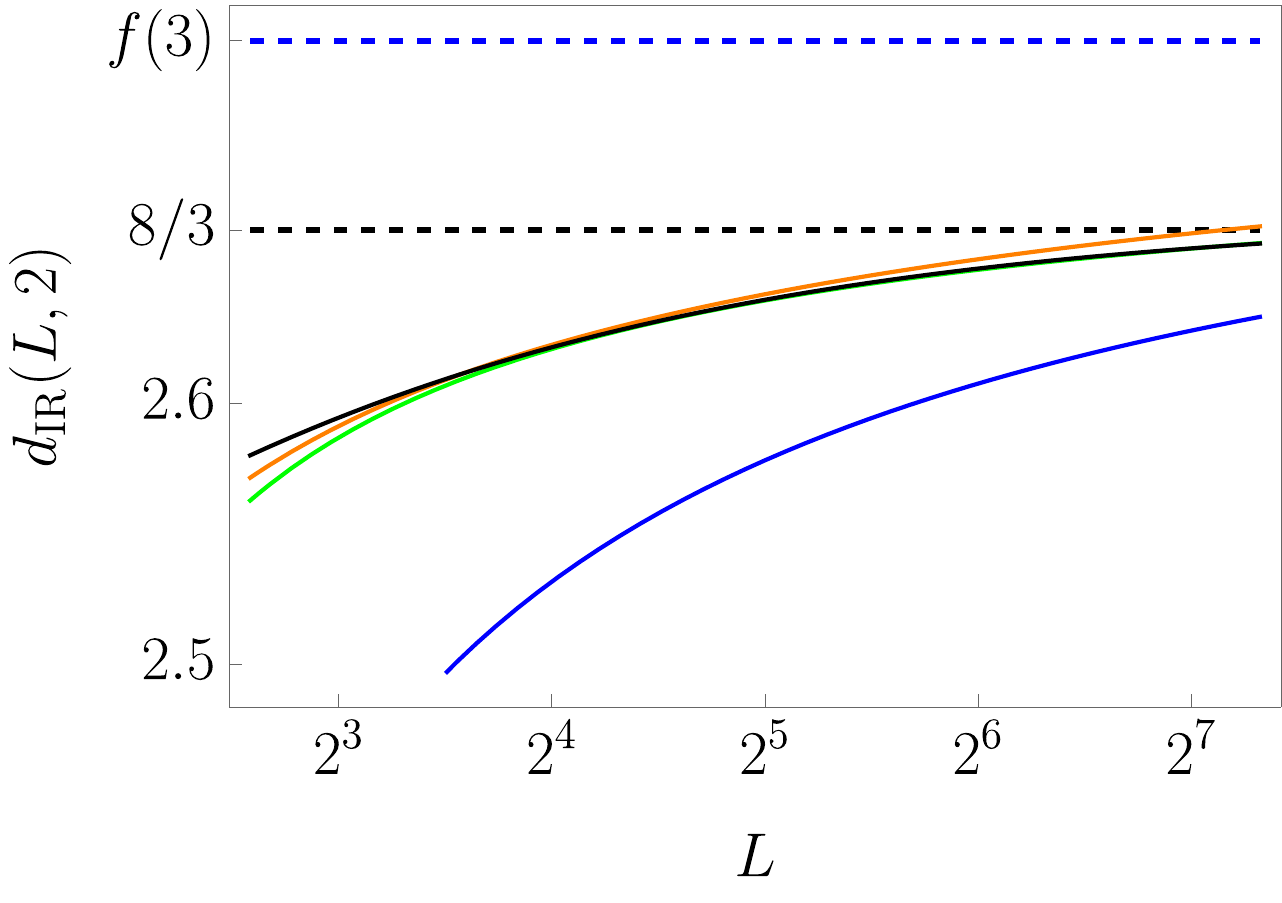}\qquad \includegraphics[width=0.95\columnwidth]{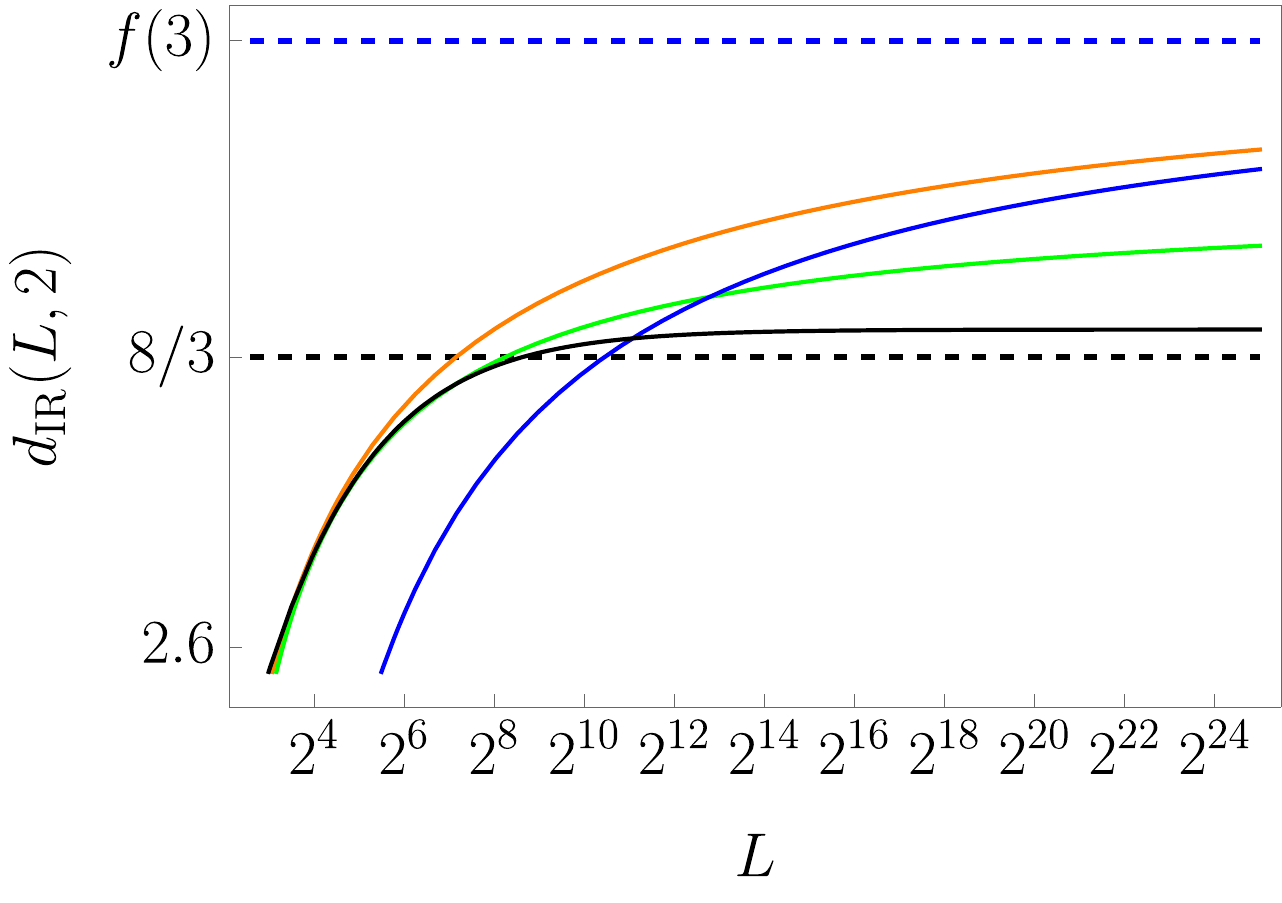}
}
\caption{Plot of 
$d_{\rm IR}(L,2)$ 
defined by Eq.~\eqref{eq:dir:def},
as a function of $L$ on the logarithmic scale. The orange curve is plotted in accordance with exact Eq. \eqref{eq:main:par} for the parabolic singularity spectrum \eqref{eq:f:p} [with $\alpha_0=4.043$ corresponding to $f(3)=2.739$]. Blue curve shows the asymptotic expression \eqref{eq:main} [with the use of the parabolic singularity spectrum \eqref{eq:f:p}]. Green curve describes the dependence given by Eq.~\eqref{eq:new:H} with $d_{\rm IR}=2.7$ and $\gamma=-0.2$. Black curve describes the dependence given by Eq.~\eqref{eq:new:H2} with $d_{\rm IR}=2.673$, $d_{\rm m}=1.998$, and $b=0.42$. ``Miraculously'', in the right panel, the orange and blue (asymptotic) curves converge. indicating the true thermodynamic behavior, while the green and the black lines strongly deviate from the exact result for the parabolic spectrum at $L\gtrsim 2^7$, thus indicating a complete failure of the data analysis employed in Ref.~\cite{Reply} for extracting the true asymptotic exponents (see text). }
\label{fig:newf}
\end{figure*}

\vspace{0.3cm}
\noindent 
\textbf{\underline{REMARK 4:} Problems with extracting $d_{\rm IR}$ from a range of small $L$.}
\color{black}
\vspace{0.2cm}

In Ref.~\cite{Reply}, the validity of asymptotic expression \eqref{eq:main} for description of data of Ref.~\cite{Horvath2022} has been questioned. We emphasize that Eq.~\eqref{eq:main} is an asymptotic expression valid at very large system sizes $L$ which are far beyond the range of data of  Ref.~\cite{Horvath2022} (see Fig. \ref{fig:2}). 
Nevertheless, the authors of Ref.~\cite{Reply} claimed that the expression
\begin{equation}
\langle \mathcal{N}_* \rangle_{\rm fit} \simeq {\rm const}\ (\ln L)^{\gamma} L^{d_{\rm IR}}
\label{eq:new:H}    
\end{equation}
with $d_{\rm IR}=2.704(1)$ and $\gamma=-0.202(2)$ provides a good fit of their data in the range $6<L<160$. Moreover, the authors of Ref.~\cite{Horvath2022} state that the good fit of their data with the help of Eq.~\eqref{eq:new:H} contradicts the multifractal predictions. This is blatantly incorrect. 

As we illustrate in Fig.~\ref{fig:newf} [for the parabolic singularity spectrum \eqref{eq:f:p}], although  Eq.~\eqref{eq:new:H} describes the exact expression \eqref{eq:main:par} for $\langle \mathcal{N}_* \rangle$ at small values of $L$, the asymptotic expression \eqref{eq:main} correctly describes behavior of $\langle \mathcal{N}_* \rangle$ from Eq.~\eqref{eq:main:par} at large values of $L$. On the contrary, expression \eqref{eq:new:H} fails to approximate the exact curve in the thermodynamic limit.

The example of an exactly solvable model of a parabolic multifractal spectrum [Eq.~\eqref{eq:f:p}], illustrated in Fig.~\ref{fig:newf}, clearly shows that the corresponding scaling function [Eq.~\eqref{eq:main:par} here] may approach the asymptotic infrared expression only at extremely large scales, which are far beyond the capabilities of numerical simulations ($L\gtrsim 16000000$ in the present example; cf.~$L\leq 160$ in Ref.~\cite{Reply}). A comparison of the two panels of Fig.~\ref{fig:newf} highlights a deceptive ``quality'' of the small-$L$ fit (green and black curves) for such type of quantities. The same deficiency also characterizes the quantity $d^u_\text{IR}(L)$ shown in Fig.~3 of Ref.~\cite{Reply}, which was also extracted by means of a simple-minded power-law fit.

The authors of Ref. \cite{Reply} argued that the ``two-power fit'' defined by expression
\begin{equation}
\langle \mathcal{N}_* \rangle_{\rm fit} \simeq {\rm const}\ (L^{d_{\rm IR}} + b L^{d_{\rm m}})
\label{eq:new:H2}    
\end{equation}
with $d_{\rm IR}=2.673(2)$ and  $d_{\rm m}=1.998(29)$ provided a good fit of their data in the same range $6<L<160$. 
For illustration purposes, in Fig.~\ref{fig:newf}, we put this fitting curve (black) on top of the curves for the parabolic singularity spectrum (as we have shown in Addendum I, those curves closely approximate the true data).
One sees that, although Eq.~\eqref{eq:new:H2} is close to the exact expression \eqref{eq:main:par} for $\langle \mathcal{N}_* \rangle$ at small values of $L$, it strongly deviates from it at large values, $L\gtrsim 2^7$. This clearly shows that the "two-power fit" of quantities that do not obey purely power-law scaling is simply not reliable when it is performed for the numerically accessible system sizes, as was done in Ref.~\cite{Horvath2022} for $\langle \mathcal{N}_* \rangle$. This is the main methodological deficiency of Ref.~\cite{Horvath2022}, which led the authors to an incorrect conclusion about the ``super-universality''. With the usage of an unjustified fitting function, the error bars in their analysis are strongly underestimated. 

The examples shown in Fig.~\ref{fig:newf} demonstrate a  well-known fact: a non-questionable fit can be done  only for data which is expected to be a straight line. We emphasize that the moments of wave functions (or the moments of local density of states) are quantities exactly of that type since they depend on $L$ as a pure power-law without any additional corrections. This is, in fact, one of the reasons behind the success of the conventional multifractal analysis performed numerically (see, e.g.,  Refs.~\cite{Rodriguez, Ujfalusi, Lindinger}). Indeed, the simple power-law scale dependence (a straight line on a log-log scale) of truly scaling observables can be then captured starting already in sufficiently small systems, yielding the true asymptotic results.   

On the contrary, the quantity $\langle \mathcal{N}_* \rangle$ analyzed in Ref. \cite{Horvath2022} is not of the ``straight-line type''. This is because it does not directly correspond to scaling eigenoperators, and thus has a complicated dependence on $L$
[cf. Eq.~\eqref{eq:main:par} and Eq.~\eqref{Nstar-logs}]. As a result, fitting this or related quantities with power-law functions (even with logarithmic corrections) yields erroneous outcomes for the asymptotic values, when the fit is performed in the range of not too large scales. Instead, one should employ the full function describing the scale dependence following from the multifractality theory.

It would be very instructive for the authors of Ref.~\cite{Reply} to repeat their analysis for the orange curve from Fig.~\ref{fig:newf} (dashed curve in Fig. 2 of Ref.~\cite{Reply}) and see  
the results for $d_\text{IR}$ 
extracted with the ``two-power fit'' \cite{Reply} performed in the range 
$L\leq 160$. Assuming a model with a strict parabolicity of the multifractal spectral function, for which the 
whole $L$-dependence of $\langle \mathcal{N}_* \rangle$ and the value of $d_\text{IR}$ are known analytically [i.e., treating Eq.~\eqref{eq:main:par} as if it is an ``experimental'' curve], such an exercise would be a good test of the methodology of Refs.~\cite{Horvath2022} and \cite{Reply} as employed to real multifractal models.

\vspace{0.3cm}
\noindent 
\textbf{\underline{REMARK 5:} On the scenario with flat region in $f(\alpha)$.}

\color{black}
\vspace{0.2cm}

In Ref. \cite{Horvath2022}, the authors suggested the singularity spectrum with a flat horizontal region (see Fig. 4 of Ref. \cite{Horvath2022}). In particular, the following form of $f(\alpha)$ has been proposed  
\begin{equation}
    f(\alpha)= 
    \begin{cases}
    \alpha, & \quad \alpha_-<\alpha<\alpha_+ ,\\
     \alpha_+, & \quad \alpha_+\leqslant \alpha \leqslant d .
     \end{cases}
     \label{eq:rel1}
\end{equation}
Invoking the symmetry relation \eqref{eq:sym}, we obtain  
\begin{equation}
    f(\alpha)= 
    \begin{cases}
    \alpha, & \quad d<\alpha<2d-\alpha_+ ,\\
     d, & \quad 2d-\alpha_+\leqslant \alpha \leqslant 2d-\alpha_- ,
     \end{cases}
     \label{eq:rel2}
\end{equation}
i.e., two more linear and flat regions in $f(\alpha)$ should exist. Moreover, 
Eqs. \eqref{eq:rel1} and \eqref{eq:rel2} together imply that $f(\alpha_+)=\alpha_+\equiv d$, meaning that the hypothetic flat region can only be at the top of the function (in fact, $f(\alpha)$ is a convex function with $f(\alpha)\leqslant d$ \cite{Evers2008}). 
However, even this scenario is impossible, because the nonanaliticity of $f(\alpha)$ at $\alpha_+$ would immediately violate the pure scaling of the moments of the wave functions, see Remarks above. 
Of course, no signs of such a drastic change of behavior was observed in the previous numerical studies of multifractality at the Anderson transition.

Thus, symmetry \eqref{eq:sym}, in combination with the exact power-law scaling of the wave-function moments, prohibits an existence of an intermediate flat region in the multifractality spectrum, which would separate two regions of growing $f(\alpha)$ (as proposed in Fig. 4 of Ref.~\cite{Reply}).

\vspace{0.3cm}
\noindent \textbf{\underline{CONCLUSIONS:}}
\vspace{0.1cm}

To summarize, the above Remarks demonstrate that 
\begin{itemize}
    \item There is no distinction between MF and mMF \cite{Reply} for Anderson transitions: $f(\alpha)\equiv f_m(\alpha)$.
    
    \item All the properties of the quantity $\langle \mathcal{N}_*\rangle$ \cite{Horvath2022} are encoded in the multifractality singularity spectrum $f(\alpha)$.
    
    \item The function $f(\alpha)$ obeys well-known exact constraints that preclude appearance of unphysical features hypothesized in Ref.~\cite{Reply}.
    
    \item The numerical analysis of the quantities like $\langle \mathcal{N}_*\rangle$ suffers from severe finite-size effects.  Extraction of $d_\text{IR}$ from the fit in the range $L\lesssim 160$ requires the knowledge of an analytical expression for $\langle \mathcal{N}_*\rangle$, since this quantity does not directly correspond to a single purely scaling operator of the underlying field theory, in contrast to moments of wave-function intensity.
    
    \item As a result, the asymptotic behavior of $\langle \mathcal{N}_*\rangle$ is achieved only at extremely large scales. Hence, any attempts of determination of $d_\text{IR}$ by means of power-law fitting at numerically accessible scales cannot be trusted. 
    
    \item The numerical results of Refs.~\cite{Horvath2022} and \cite{Reply} are in an exceptionally good agreement with the multifractality theory of Anderson transitions (cf. Fig.~\ref{fig:2}), when the data is treated in a proper way, using the known analytical results. 
\end{itemize}
Thus, the main conclusion of the initial version of the Comment remains undeniable after the Responses \cite{Reply}:  $d_{\rm IR}$ (introduced in Ref.~\cite{Horvath2022}) \textbf{is nothing but} $f(d)$.

\end{document}